\documentclass[a4paper]{article}
\usepackage{icrc2013}

\newcommand{\g}{$\gamma$}

\newcommand{\degr}{$^{\circ}$}                         

\title{The 1st {\emph Fermi} LAT SNR Catalog: Probing the Gamma-ray Population}

\shorttitle{{\emph Fermi} LAT SNR Catalog}

\authors{
J.W. Hewitt$^{1,2}$,
F. Acero$^{2}$,
T.J. Brandt$^{2}$,
J. Cohen$^{3}$
F. de Palma$^{4}$,
F. Giordano$^{4}$
for the {\emph Fermi} LAT Collaboration.
}

\afiliations{
$^1$ CRESST/University of Maryland, Baltimore County, Baltimore, MD 21250, USA \\
$^2$ NASA Goddard Space Flight Center, Greenbelt, MD 20771, USA\\
$^3$ Department of Physics and Department of Astronomy, University of Maryland, College Park, MD 20742, USA \\
$^4$ INFN Sezione di Bari, 70126 Italia\\
}

\email{john.w.hewitt@nasa.gov}

\abstract{
While supernova remnants (SNRs) are widely thought to be powerful accelerators, evidence
comes largely from a small number of well-studied cases. Here we systematically determine
the \g-ray emission from all known Galactic SNRs, disentangling them from the sea of
Galactic cosmic rays. Using {\emph Fermi} LAT data we have characterized the GeV emission in all
regions containing SNRs, accounting for systematic uncertainties caused by source
confusion, diffuse emission modeling, and instrumental response.  More than a dozen
remnants are identified through spatial extension or detection at TeV energies, with
potential associations for $>$40 more. From this population study, two clear classes of
\g-ray-emitting SNRs emerge: young remnants and those interacting with a dense medium.
This large statistical sample also reveals a possible correlation between GeV and radio
flux. The growing number of identified SNRs will help to disentangle the effects of age
and environment on the aggregate properties of SNRs at high energies.
}

\begin{document}
\maketitle

\section{Introduction}
A key question that \g-ray astronomy seeks to answer is the origin of Galactic cosmic rays. The current generation of ground- and space-based \g-ray observatories allows for detecting the extension of numerous suspected accelerators, including SNRs. 
Emission from SNRs in the GeV energy range gives an important window into the emission mechanism, and total energetics of accelerated particles. Previous searches of bright radio SNRs have revealed a small number of possible counterparts \cite{esposito1996}, but only with the {\emph Fermi} LAT telescope has the sensitivity been obtained over the full-sky to detect a significant fraction of Galactic SNRs. 

{\emph Fermi} LAT has identified a number of SNRs (see \cite{thompson2012} for a review), but their emission has been largely explored on a source-by-source basis. The near-uniform sky coverage allows a systematic exploration of both detections and non-detections of the hundreds of known SNRs in our Galaxy. Here we present our ongoing effort to systematically explore \g-ray emission from SNRs for the First {\emph Fermi} LAT Supernova Remnant Catalog. We broadly detail the analysis pipeline (\S 2), discuss the classification of detected sources (\S 3) and highlight one interesting aspect of the catalog, a direct comparison between GeV and radio emission (\S 4).

\section{Detection Pipeline Description}

\subsection{Gamma-ray Data}

The \emph{Fermi} LAT is a pair-conversion \g-ray telescope detecting photons from 20~MeV to $>300$~GeV \cite{Atw09}. 
Our catalog is constructed from 3 years of LAT survey data and the Pass7v6 instrument response functions (IRFs). For each of the 278 SNRs identified in Green's catalog \cite{green2009} 
we modeled emission within a 10\degr\ radius of interest (ROI) from the SNR center. 
As a compromise between sensitivity, spatial resolution to resolve extension and to separate the SNR from the diffuse background, we choose 1 GeV as our minimum energy threshold. Only source class events are selected with energies of 1 to 100 GeV.

To analyze the data we fit models using the maximum likelihood framework. Tools utilized include both the standard science tools\footnote{Available at the {\emph Fermi} Science Support Center: \url{http://fermi.gsfc.nasa.gov/ssc}} and the {\tt pointlike} analysis package \cite{kerr2010} which has been specifically developed and verified for characterizing source-extension for {\emph Fermi} LAT data \cite{lande2012}.

\subsection{Initial Source Model}

In order to characterize each SNR, we must obtain an optimal characterization of \g-ray emission in the ROI that includes all significant sources of emission. To do so, we developed an automated analysis pipeline briefly described here. We start from the standard models of diffuse emission and a list of identified sources in the Second {\emph Fermi} LAT catalog (2FGL) \cite{nolan2012}. Using {\tt pointlike} we generate a map of source test statistic (TS) for each 0.1\degr\ bin covering the entire ROI. Here the source TS is defined as twice the logarithm of the ratio between the likelihood $\mathcal{L}_1$ obtained by fitting the source model plus background components (including other sources) to the data, and the likelihood $\mathcal{L}_0$ obtained by fitting the background components only, i.e., TS = 2 log($\mathcal{L}_1$/$\mathcal{L}_0$). At the position of the peak TS value we add a new point source with a power-law spectral model, perform a likelihood fit of the region, and localize the position of the newly added source (only the first time it is added). 
This iterative process is continued within the specified region of interest until there are no remaining sources which change in the global log-likelihood by more than 8.
This threshold leads to the detection of all TS $\geq$ 25 sources. Our final step is a removal of all sources with TS $<$ 25 from the final model.

\subsection{Source Localization and Extension}

\begin{figure}[t]
  \centering
  \includegraphics[width=0.45\textwidth]{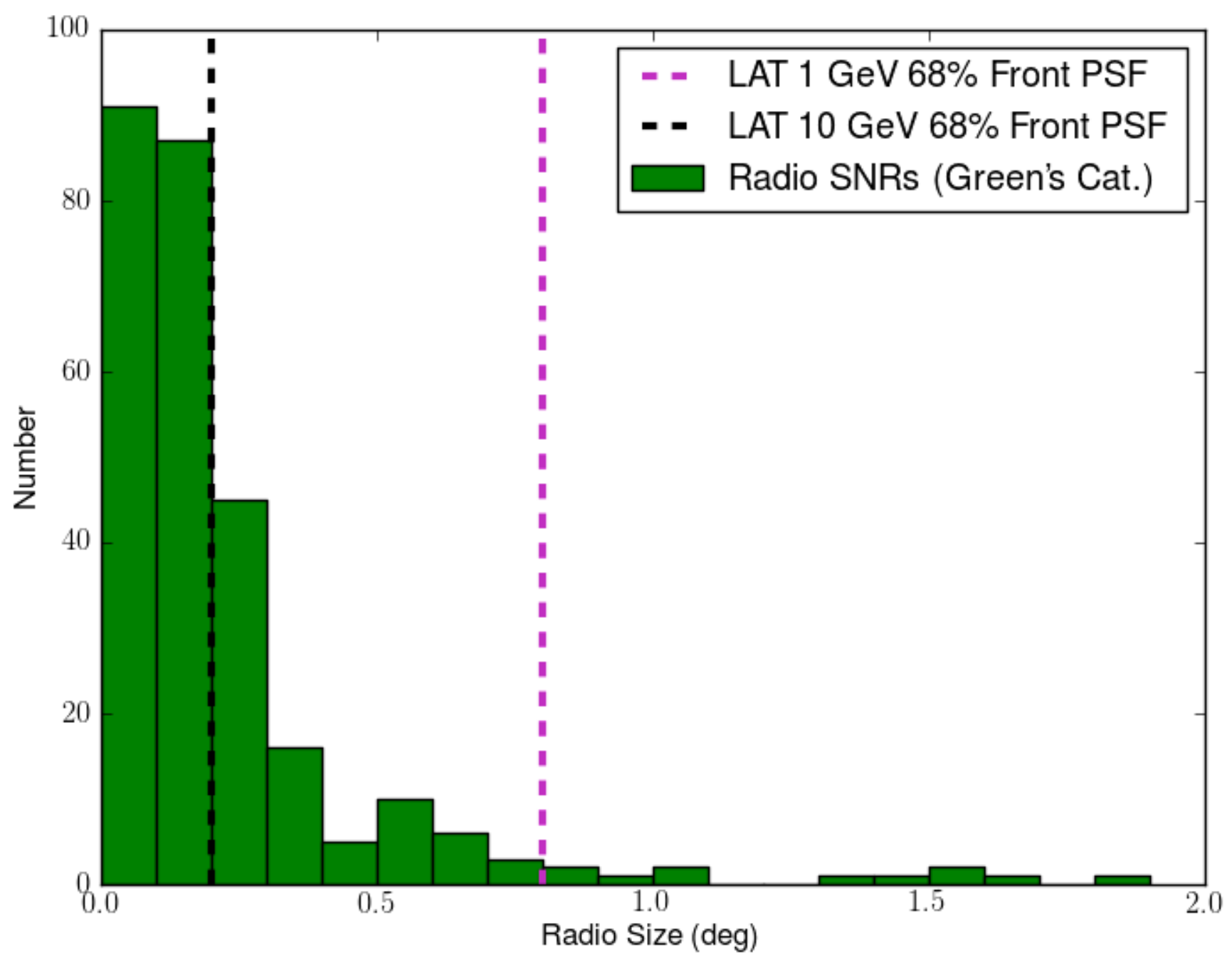}  
  \caption{Comparison of the radio diameter in Green's catalog. The vertical dashed lines indicates the 68\% containment radius of the LAT at 1 and 10 GeV for front-converting events, which is roughly equivalent to the limit at which bright sources can have detectable extensions.}
  \label{fig:size_hist_p}
\end{figure}

Many \g-ray sources are detected coincident with the position of known SNRs, however this is not sufficient to make an identification. The detection of spatial extension remains the best way to determine that \g-ray emission originates from the SNR. The spatial resolution of the LAT is sufficient to detect many SNRs as extended. Figure \ref{fig:size_hist_p} shows the distribution of radio sizes from Green's catalog. The 68\% containment radius at 1 and 10 GeV are indicated as vertical dashed lines that roughly approximate the threshold for the detection of SNRs if they are sufficient bright \g-ray sources. Roughly a third of all SNRs may potentially be resolved by the LAT.

For each SNR we use our analysis pipeline to characterize the morphology and spectrum of any \g-ray emission that may be coincident with the SNR location (as defined by the position and extent in the radio reported in Green's catalog). All sources which fall within the radio radius are removed from the model, unless the source has been previously identified as not an SNR (e.g. pulsars).

Several hypotheses are then explored in parallel using {\tt pointlike}. For the point source hypothesis, a point source is placed at the radio centroid of the SNR. Sources within 5\degr\ of the SNR center are fit with the normalization left free but the spectral index fixed. For the disk hypothesis, a uniform disk equal in radius to the radio size is placed at the SNR center. The disk normalization, index, position and extension are fit. In a separate hypothesis, we will also test the significance of sources which are adjacent to the SNR disk. 
We will determine whether a nearby source is kept or removed from the final model by defining TS$_{\rm nearby}$ as twice the difference between the model with the nearby source and the model without the source, in which the extension and position of the disk are refit. A nearby source is significant (and thus kept in the final model) if TS$_{\rm nearby}$ $\geq$ 9. 

Once these hypotheses have been evaluated, we compare the global log-likelihoods of all the resulting models to determine which model gives the most significant representation of the data. A source is considered possibly associated with an SNR if a point source with TS $>$ 25 is detected within the radio extent of the SNR. To determine whether a source has a significant extension we define TS$_{ext}$ as twice the difference between the log-likelihood of the final model from the disk hypothesis and that of the point-source hypothesis. 
A SNR is considered to have a significant extension if TS$_{ext}$ $\geq$ 16 as in \cite{lande2012}. 
The detection of GeV extension in agreement with the measured extension of the remnant at other wavelengths provides a secure means of identifying the SNR in \g-rays.

\subsection{Fluxes and Upper Limits}\label{sec:Spec}

Fluxes are determined in the 1-100 GeV band given the best, significant characterization of the morphology and spectra of the SNR from the pipeline detailed above. For those SNRs which are not significantly detected we compute upper limits on the flux by assuming a spectral model with a power-law index of $2.5$, consistent with the majority of SNRs detected. As a spatial model we use the uniform disk equal in position and radius to that reported in Green's catalog. We then calculate an upper limit on the flux from the disk without including any overlapping sources in the model which have not been firmly identified.

\subsection{Sources of Error}

In addition to statistical error, we will account for several sources of systematics in our catalog. The two main sources of systematics are uncertainties in the model of Galactic diffuse emission, and in the effective area calibration. Uncertainties in the effective areas are estimated using modified IRFs that bracket the nominal ones. Characterizing uncertainties in the diffuse model required the construction of alternative diffuse models, and is detailed in \cite{deplama2013}. Detections reported in the final catalog will be required to be robust against these sources of systematic error.

\section{SNR Identification and Emerging Classes}

Using the pipeline described above, we detected 44 SNRs at energies of $>$1 GeV. Of these, 15 are extended (with 6 being new detections), 4 are spatially unresolved SNRs which we identify based on TeV detections, and 25 are candidate associations which do not show significant extension. This brings the total number of identified SNRs at GeV energies to 19, which is sufficient to begin to explore properties of the population.

Among the point-like SNRs which cannot be firmly identified, we attempt to qualitatively classify whether the GeV source is more likely to be of pulsar- or SNR-origin. We examined archival X-ray data to search for evidence of a possible point source counterpart. Those which do not contain a known pulsar (in radio or X-rays) and do not have an identified X-ray point source within their interiors we classify as favorable SNR-candidates, while the others are denoted as unfavorable SNR-candidates. We note that this classification is only a qualitative assessment, useful to examine whether differences are apparent between those point-like SNR candidate associations.

\begin{figure}[!t]
  \centering
  \includegraphics[width=0.45\textwidth]{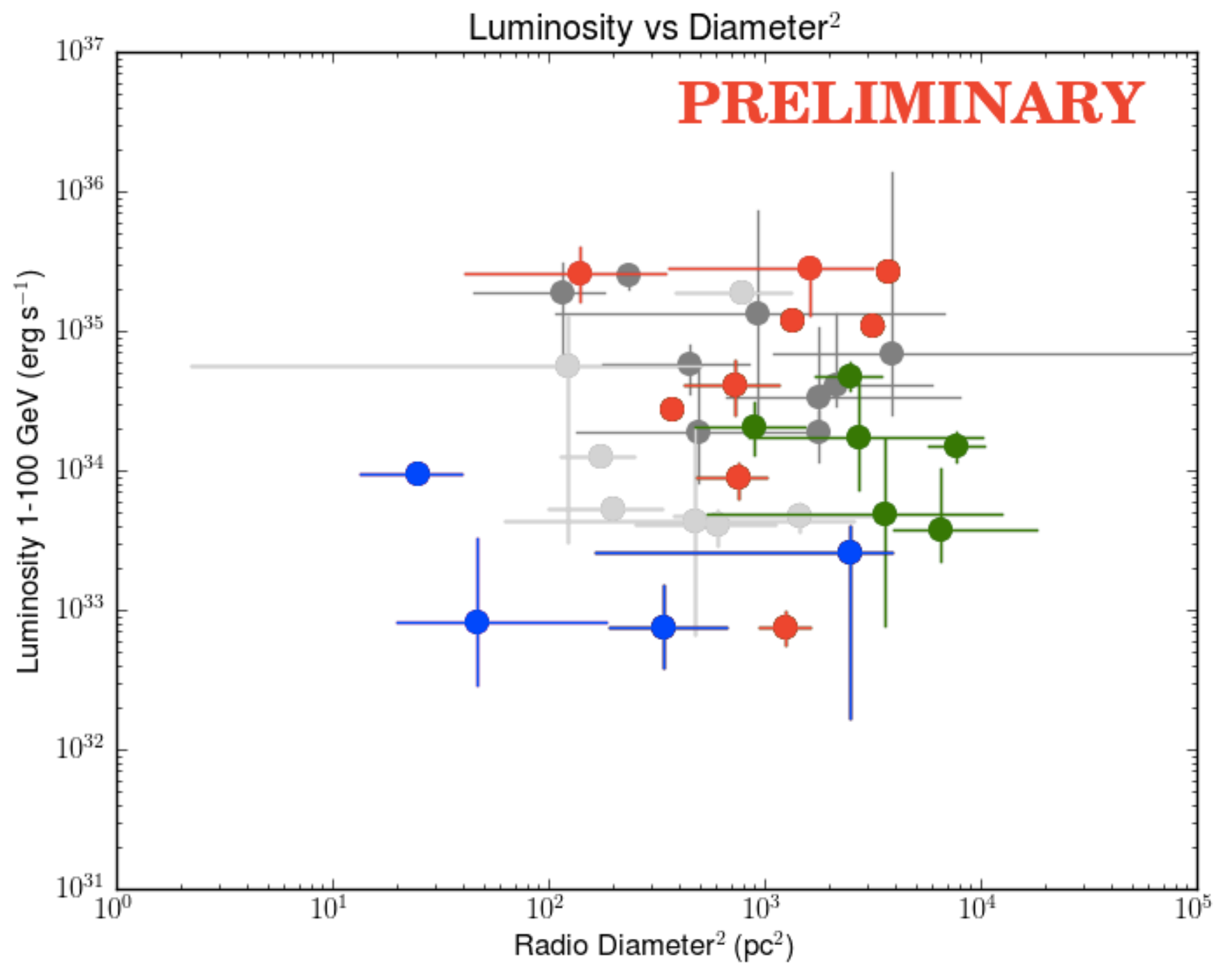}
  \caption{The luminosity of detected SNRs at 1--100 GeV energies plotted as a function of the radio diameter squared, which is a function of both the age of the SNR and the density of the environment into which it is expanding. Error bars include statistical uncertainties from fitting the GeV data as well as errors in the published distances. We use colors to indicate this classification as follows: identified young SNRs in blue, identified interacting SNRs in red, newly identified SNRs in green, favorable candidate SNRs in dark grey, and unfavorable candidate SNRs in light grey.
}
  \label{fig:lum-dia}
 \end{figure}

From our sample of GeV-identified SNRs, two clear classes emerge. The largest class is those SNRs known to be interacting with molecular clouds, which are typically quite luminous at GeV energies. In contrast, the few known young SNRs are less luminous, with harder spectra and TeV-detections. These two classes are clearly separated in Figure \ref{fig:lum-dia} which plots the 1--100 GeV luminosity against the physical diameter squared, which acts as a proxy for the evolutionary state of the SNR (more evolved SNRs will have larger diameters). Some SNRs are not included because they do not yet have distance estimates. Detected SNR luminosities span more than two orders of magnitude. The newly identified SNRs appear as a lower luminosity extension of the interacting SNR class. 

\section{GeV--Radio Correlation}

\begin{figure}[!t]
  \centering
  \includegraphics[width=0.45\textwidth]{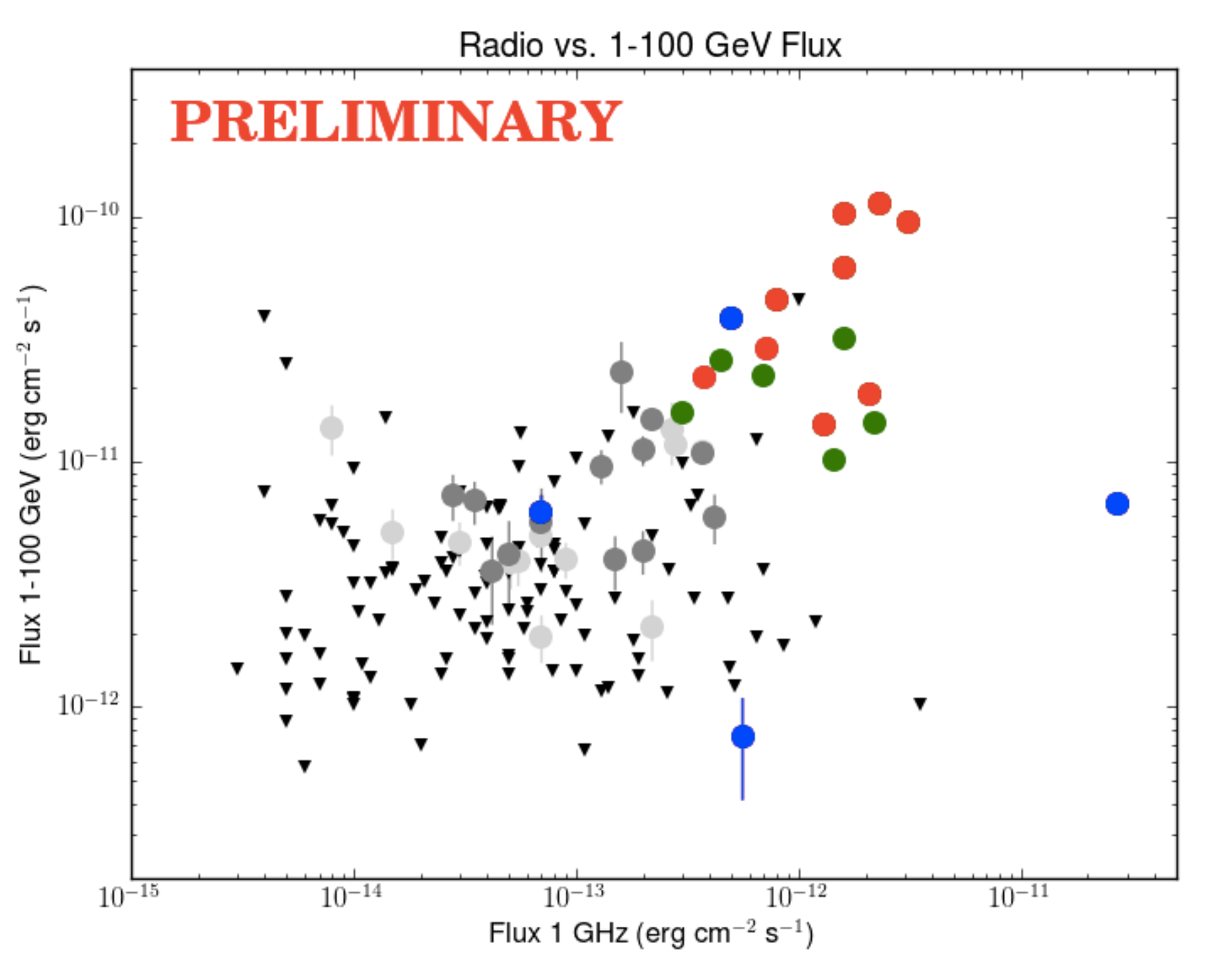}
  \caption{Comparison of Radio vs. GeV flux including upper limits. See the caption of Fig. \ref{fig:lum-dia} for a description of colors.}
  \label{fig:gev-radio}
 \end{figure}

A correlation between the GeV and radio flux from SNRs may be expected, as both result from nonthermal emission of relativistic particles. It has been noted that interacting SNRs are preferentially radio-bright \cite{hewitt09}. The presence of a large target mass for recently accelerated or escaping cosmic rays by the SNR would increase the \g-ray emission via $\pi^0$-decay or electron bremsstrahlung emission. Figure \ref{fig:gev-radio} compares the 1 GHz radio flux density to the 1--100 GeV photon flux for all SNRs in our catalog. We include upper limits for those SNRs which do not have any coincident GeV emission detected. For most identified SNRs (red, green) and those which we deem more likely to be SNRs (dark grey) a clear trend is apparent. Young SNRs do not appear to follow this trend, perhaps indicating a different emission mechanism. We note that some SNRs which fall below this correlation, having fainter GeV fluxes than expected given their radio flux density, also appear to have softer GeV power-law indices. Many SNRs detected by {\emph Fermi} LAT at GeV energies are not detected at TeV-energies due to spectral curvature around GeV energies. Deviations from this correlation may reflect processes related to the acceleration and escape of cosmic rays from the SNRs.

\section{Summary}

We present a systematic survey of GeV emission from all known Galactic SNRs using 3-years of data from {\emph Fermi} LAT. Our automated pipeline characterizes emission in all regions containing SNRs in Green's catalog, accounting for systematic uncertainties caused by source confusion, diffuse emission modeling, and instrumental response. We have identified 19 source as SNRs and 25 sources as possible associations. From this population of detected remnants, we can clearly distinguish a dominant class of SNRs interacting with a dense medium, and a less numerous class of young SNRs. The large sample of SNRs reveals a possible correlation between GeV and radio flux. The growing number of identified SNRs promises to help disentangle the mechanism of \g-ray emission from SNRs and thereby the energetics transferred into the acceleration of cosmic rays.

\vspace*{0.5cm}
\footnotesize{{\bf Acknowledgment:}{
The {\emph Fermi} LAT Collaboration acknowledges support
from a number of agencies and institutes for both development and the operation of the LAT as well as scientific data analysis. These include NASA and DOE in the United States, CEA/Irfu and IN2P3/CNRS in France, ASI
and INFN in Italy, MEXT, KEK, and JAXA in Japan, and
the K. A. Wallenberg Foundation, the Swedish Research
Council and the National Space Board in Sweden. Additional support from INAF in Italy and CNES in France for
science analysis during the operations phase is also gratefully acknowledged.
}}

\end{document}